\documentclass[aps,prb,twocolumn,groupedaddress,showpacs,showkeys,amsmath,superscriptaddress]{revtex4-1}
\usepackage[dvips]{graphicx}
\usepackage{dcolumn}
\usepackage{color}
\usepackage{colortbl}
\usepackage{longtable}
\usepackage{ctable}
\usepackage{rotating}
\usepackage{booktabs}

\begin{document}

\title{An efficient method for grand-canonical twist averaging in quantum Monte Carlo calculations} 

\author{Sam Azadi}
\email{sam.azadi@kcl.ac.uk}

\affiliation{Department of Physics, King's College London, Strand,
  London WC2R 2LS, United Kingdom}
\affiliation{Department of Physics, Imperial College London, South
  Kensington Campus, London SW7 2AZ, United Kingdom}

\author{W. M. C. Foulkes}

\affiliation{Department of Physics, Imperial College London, South
  Kensington Campus, London SW7 2AZ, United Kingdom}

\date{\today}

\begin{abstract}
  We introduce a simple but efficient method for grand-canonical twist
  averaging in quantum Monte Carlo calculations. By evaluating the
  thermodynamic grand potential instead of the ground state total
  energy, we greatly reduce the sampling errors caused by
  twist-dependent fluctuations in the particle number. We apply this
  method to the electron gas and to metallic lithium, aluminum, and
  solid atomic hydrogen. We show that, even when using a small number of
  twists, grand-canonical twist averaging of the grand potential
  produces better estimates of ground state energies than the widely
  used canonical twist-averaging approach.
\end{abstract}

\maketitle

\section{Introduction}

Many-body wave function based Quantum Monte Carlo (QMC) techniques such
as variational Monte Carlo, diffusion Monte Carlo (DMC) and auxiliary
field Monte Carlo are widely used to calculate ground- and excited-state
properties of real
materials.\cite{Matthew1,Kolorenc,Dubeck,Morales,Wagner,
  Sandro_b,Booth,Zen, Sandro11, Seki, Motta} Many materials and
properties that cannot be described accurately using single-particle
based approaches have been studied successfully using QMC methods. For
example, QMC techniques have been used to elucidate the nature of
noncovalent and weak van der Waals interactions, which are crucial in
chemistry, biology, and biochemistry.\cite{Dubecky} The most important
contribution of QMC to materials science and electronic structure theory
has perhaps been to provide input to mean-field based methods, most
notably via the QMC calculations of the homogeneous electron
gas\cite{Ceperley80} that led to the first accurate local density
approximation and have directly or indirectly contributed to almost
every exchange-correlation functional devised since then.
  
QMC calculations of the properties of crystals and solids use finite
simulation cells subject to periodic boundary conditions. The
volume of the simulation cell is strongly restricted for computational
reasons and the finite-size errors caused by the replacement of an
infinite solid by a small simulation cell are large. Controlling these
errors is one of the main challenges faced in all QMC simulations of
extended systems.\cite{Fraser,Chiesa,Drummond,Holzmann,samFS}

Within the Born-Oppenheimer approximation, the Hamiltonian of an
$N$-electron simulation cell can be expressed as
$\hat{H} = \hat{T} + \hat{V}$, where $\hat{T}$ is the electronic
kinetic-energy (KE) operator and $\hat{V}$ is the operator for the
interaction energy, including electron-electron and electron-nuclear
contributions: $\hat{V} = \hat{V}_{\text{e-e}} + \hat{V}_{\text{e-n}}$.
The expectation value of $\hat{V}_{\text{e-e}}$ is often written as the
sum of two terms:
$\langle \hat{V}_{\text{e-e}} \rangle = E_{\text{H}} + E_{xc}$. The
Hartree energy, $E_{\text{H}}$, is the classical Coulomb interaction
energy associated with the electronic charge density
$\rho_{\text{e}}(\mathbf{r})$. The exchange-correlation energy,
$E_{xc}$, contains the rest of electron-electron interaction energy,
including contributions from the correlations between the positions of
electrons and the anti-symmetry of the fermionic many-electron
wave function. The electron-nuclear interaction energy
$\langle \hat{V}_{\text{e-n}} \rangle$ and the Hartree energy
$E_{\text{H}}$ are functionals of the electronic charge density
$\rho_{\text{e}}(\mathbf{r})$, which normally converges rapidly as the
number of unit cells within the simulation cell increases. Thus, the
finite-size errors in these terms are small compared to those in other
components of the total energy. By contrast, the finite-size errors in
the exchange-correlation energy and the KE can be very substantial. In
this work, we introduce an efficient and practical method for correcting
the finite-size errors in the dominant one-electron contribution to the
KE.


\section{Single-particle finite-size problem}\label{SPFS}

In mean-field-like approaches such as Density Functional Theory (DFT),
exact results for infinite periodic crystals can be obtained by solving
the Schr\"{o}dinger equation within a single primitive unit cell subject
to Bloch boundary conditions. Expectation values per unit cell of the
infinite periodic system are obtained by integrating over the first
Brillouin zone (BZ), which is equivalent averaging over all possible
Bloch boundary conditions.

This approach does not yield exact results in many-particle methods such
as QMC. The problem is that the range of the correlations between
electron positions often exceeds the size of the primitive unit cell.
Reducing the system to one primitive cell is then no longer acceptable.
QMC simulations are instead carried out in simulation cells comprising
several primitive cell. Exact results are obtained only in the limit as
the size of the simulation cell tends to infinity.

The long-ranged many-body correlation effects are included in an
approximate way in local and semi-local DFT calculations, where they are
built in to the exchange-correlation functional. This functional,
however, was parameterized with the help of QMC simulations of large
simulation cells.

Generally, in QMC calculations of periodic systems, the Hamiltonian
$\hat{H}$ of the $N$-electron simulation cell exhibits two types of
periodicity:\cite{Rajagopal}
\begin{equation}
  \hat{H} ({\bf r}_{1}, \ldots, {\bf r}_{i}, \ldots, {\bf r}_{N}) =
  \hat{H}({\bf r}_{1}, \ldots, {\bf r}_{i} + {\bf R}_{s}, \ldots, {\bf r}_{N})
\label{eq1}
\end{equation}
for all $1 \leq i \leq N$, and
\begin{equation}
  \hat{H} ({\bf r}_{1}, {\bf r}_{2}, ..., {\bf r}_{N}) =
  \hat{H}({\bf r}_{1}+{\bf R}_{p}, {\bf r}_{2} + {\bf R}_{p}, ..., {\bf
    r}_{N} + {\bf R}_{p}),
\label{eq2}
\end{equation}
where ${\bf R}_{s}$ and ${\bf R}_{p}$ are the simulation-cell and
primitive-cell lattice vectors, and
$({\bf r}_{1},{\bf r}_{2},\ldots,{\bf r}_{N})$ are the electron
coordinates. The simulation-cell periodicity, Eq.~(\ref{eq1}), arises
from the periodic boundary conditions applied across the finite
simulation cell and does not hold in a real solid; the primitive-cell
periodicity, Eq.~(\ref{eq2}), also holds in real systems as long as
periodic boundary conditions are applied to the solid as a whole.

Because of the two types of periodicity, the $N$-electron wave function
of the simulation cell obeys two types of Bloch theorem:
\begin{equation}
  \Psi_{{\bf k}_s} = V_{{\bf k}_{s}} ({\bf r}_{i}, \ldots ,{\bf r}_{N})
  \exp \left ( i{\bf k}_s \cdot \sum_{i=1}^{N} {\bf r}_i \right ) ,
\label{eq3}
\end{equation}
\begin{equation}
\Psi_{{\bf k}_p} = U_{{\bf k}_{p}} ({\bf r}_{i},...,{\bf r}_{N}) \exp
\left ( i {\bf k}_p \cdot \frac{1}{N}\sum_{i=1}^{N} {\bf r}_i \right ) ,
\label{eq3}
\end{equation}
where $V_{{\bf k}_s}$ is invariant under the translation of any one
electron by a simulation-cell lattice vector ${\bf R}_s$ and
$U_{{\bf k}_p}$ is invariant under the simultaneous translation of all
$N$ electrons by a primitive lattice vector ${\bf R}_p$. Without loss of
generality, we can assume that the simulation-cell wavevector
${\bf k}_s$ lies within the simulation-cell Brillouin zone and that the
primitive Bloch wavevector ${\bf k}_p$ lies within the primitive
Brillouin zone (which is, of course, larger).

A many-body simulation with a non-zero ${\bf k}_s$ is normally described
as subject to twisted boundary conditions,\cite{Lin} and averaging the
results over different twists is called twist averaging. The technique
of twist averaging can be carried out in the canonical ensemble (CE),
which fixes the number of electrons in the simulation cell, or in the
grand canonical ensemble (GCE), which allows the number of electrons to
vary with the twist ${\bf k}_s$. Because of the existence of a sharp
Fermi surface and shell-filling effects, the use of twisted boundary
conditions (TBC) is more important in metals than in
insulators.\cite{Lin}

To clarify the origin of the shell-filling effects, consider a finite
simulation cell of non-interacting electrons subject to twisted boundary
conditions. The one-electron potential has the periodicity of the
primitive unit cell, so the single-particle orbitals adopt the usual
Bloch form, $\psi_{\bf k}=u_{\bf k}({\bf r}) \exp(i{\bf k.r})$, where
$u_{\bf k}({\bf r})$ has the periodicity of the primitive cell. The
twisted boundary conditions require the Bloch wavevector to lie on a
grid of points of the form ${\bf k} = {\bf k}_s + {\bf G}_s$, where
${\bf G}_s$ is a reciprocal vector of the simulation-cell lattice. There
are exactly $N_c$ such reciprocal vectors within the primitive Brillouin
zone, where $N_c$ is the number of primitive unit cells in the
simulation cell. To make this more concrete, consider a simulation cell
consisting of $N_c = L \times L \times L$ primitive unit cells. The
Bloch wavevectors then lie on an $L \times L \times L$ Monkhorst-Pack
grid \cite{MPgrid} within the primitive Brillouin zone, offset from the
origin by the twist ${\bf k}_s$, which lies within the simulation-cell
Brillouin zone.

To calculate, for example, the total non-interacting KE at twist
${\bf k}_s$, a sum over contributions from the occupied orbitals at all
$N_c$ distinct ${\bf k}$ points of the form ${\bf k}_s + {\bf G}_s$ is
carried out. In an infinite simulation cell, the sum becomes an integral
over the Brillouin zone, including contributions from every
single-particle orbital below the Fermi energy $E_f$. The grid of
simulation-cell reciprocal lattice vectors ${\bf G}_s$ becomes finer as
the size of the simulation cell increases, so the integrand is sampled
more finely for larger simulation cells. In an insulator, where the
integrand is a smooth function of ${\bf k}$, a coarse quadrature grid is
sufficient to yield accurate results; but in metals, where the bands
that cross the Fermi level are occupied in some parts of the Brillouin
zone and unoccupied in others, the integrand is discontinuous and the
quadrature errors are large. It is then necessary to increase the size
of the simulation cell or average over more twists to obtain accurate
results.

In non-interacting systems, these two approaches (increasing the size of
the simulation cell or averaging over more twists) are equivalent and
both are capable of giving exact results. In interacting systems,
increasing the size of the simulation cell still gives exact results,
but averaging over twists applied to a finite simulation cell does not.
Because of the long-ranged electronic correlations, no many-body
simulation for a finite simulation cell can be exact. In practice, we
make the simulation cell as large as computational limitations allow and
twist average to reduce the single-particle contributions to the size
error. The residual many-body size errors, which are not removed by
twist averaging results for a finite simulation cell, are tackled using
other methods.\cite{Fraser,Chiesa,Drummond,Holzmann}

In QMC simulations of spin-unpolarised systems, the canonical
twist-averaging approach works as follows. For every twist ${\bf k}_s$,
one constructs the determinantal part of the QMC trial wave function by
collecting the one-electron orbitals (usually obtained from a DFT or
Hartree-Fock calculation) associated with all $N_c$ points on the
quadrature grid of points of the form ${\bf k} = {\bf k}_s + {\bf G}_s$.
The $N/2$ orbitals of lowest energy are then doubly occupied. This
guarantees that the number of electrons in the simulation cell is
independent of twist ${\bf k}_s$ and always equal to $N$. In the
grand-canonical twist-averaging approach, only those one-electron
orbitals for which the mean-field (DFT or Hartree-Fock) energy
eigenvalue lies below the mean-field Fermi energy are doubly occupied.
Hence, the number of electrons depends on ${\bf k}_s$.

In non-interacting systems, grand-canonical twist averaging is exactly
equivalent to conventional Brillouin zone integration, which also
considers contributions only from orbitals within the Fermi surface. As
the number of twists tends to infinity, exact results are obtained.
Canonical twist averaging occasionally occupies orbitals outside the
Fermi surface and occasionally leaves orbitals within the Fermi surface
unoccupied. Assuming that the curvature of the bands crossing the Fermi
energy is positive, this add a small positive bias to the energy
estimate, even in a non-interacting system.

\section{Grand-canonical twist averaging of the grand potential}
\label{new_app}

The conventional grand-canonical twist-averaging method is not generally
viewed as a practical approach because of the strong sensitivity of the
total energy of the simulation cell to the twist. This is primarily due
to the ${\bf k}_s$-dependence of the number of electrons within the
simulation cell. It is difficult to get accurate results without
sampling impractically large numbers of twists.\cite{Drummond,Holzmann}

To reduce the cost of twist averaging in the CE, various techniques
based on the selection of optimal twists have been
introduced.\cite{Rajagopal,Dagrada,Mihm} In this section, we introduce a
new approach to twist averaging in the GCE, allowing total, kinetic,
exchange, and correlation energies to be obtained accurately without
using very many twists. The uncertainties in results obtained using the
new GCE twist-averaging algorithm are comparable to those in CE
twist-averaging calculations. Unlike CE twist averaging, however, GCE
twist averaging removes independent-particle finite-size errors exactly
as the number of twists tends to infinity, even for small simulation
cells. GCE twist averaging is thus in general to be preferred to CE
twist averaging. A similar approach has previously been used to control
the finite-size errors in exact diagonalization studies of the one- and
two-dimensional Hubbard model,\cite{Gros} and Ref.~\onlinecite{Dagrada}
suggests the use of a similar technique in QMC, but we are not aware of
previous applications to continuum QMC simulations.

In the conventional GCE twist-averaging approach, results are obtained
by twist averaging the total energy,
\begin{equation}
  \label{eq:Eave}
  E = \frac{1}{M} \sum_{{\bf k}_s} E({\bf k}_s) ,
\end{equation}
where the sum is over the sample of $M$ twist vectors
${\bf k}_s$ and $E({\bf k}_s)$ is the total energy for twist
${\bf k}_s$. If we consider a Hartree-Fock calculation with only a
single band for simplicity, $E({\bf k}_s)$ is the energy of the Slater
determinant containing all one-electron orbitals
$\psi_{{\bf k}_s + {\bf G}_s}({\bf r})$ with ${\bf k}_s$ fixed and
${\bf G}_s$ chosen such that $|{\bf k}_s + {\bf G}_s|$ lies within the
Fermi surface.

Energies obtained using Eq.\ (\ref{eq:Eave}) are inaccurate for small
numbers of twists because the number of orbitals in the Slater
determinant is surprisingly sensitive to the twist ${\bf k}_s$. If, for
example, we consider a uniform electron gas with
$r_s$$=$1, choosing the system size such that the face-centered-cubic
(FCC) simulation cell contains 118 electrons on average, the actual
electron number varies from 102 to 128 (at least) as ${\bf
  k}_s$ varies. These $\pm
10$\% fluctuations in particle number yield similarly large fluctuations
in the values of $E({\bf
  k}_s)$ and hence slow convergence of the mean
$E$ with the number of twists.

The observation that leads to a better algorithm is that the
thermodynamic free energy appropriate for use with the grand-canonical
ensemble is not the internal energy $E$ but the grand potential
\begin{equation}
  \Omega(T, V, \mu) = E(S, V, N) - TS - \mu N ,
\end{equation}
where the entropy $S$ and particle number
$N$ appearing on the right-hand side are to be regarded as functions of
the temperature $T$, the volume $V$, and the chemical potential
$\mu$.  Since we are working at zero temperature and fixed volume, we
simplify this to
\begin{equation}
  \label{eq:legendre}
  \Omega(\mu) = E(N) - \mu N .
\end{equation}

The clearest way to formulate the Legendre transformation that yields
$\Omega(\mu)$ from $E(N)$ is to start with a function of \emph{two}
independent variables, $\mu$ and $N$,
\begin{equation}
  \Omega(\mu, N) = E(N) - \mu N ,
\end{equation}
and define $\Omega(\mu)$ via a minimisation:
\begin{equation}
  \Omega(\mu) = \text{Min}_{N} \Omega(\mu, N) =
  \text{Min}_{N} \left ( E(N) - \mu N \right ) .
\end{equation}
This variational definition shows explicitly that the free energy
$\Omega$ is a function of $\mu$, not $N$, and yields, if we treat $N$ as
continuous, the minimisation condition,
\begin{equation}
  \label{eq:Nofmu}
  \frac{d E}{dN} = \mu ,
\end{equation}
from which one obtains the function $N(\mu)$ appearing on the right-hand
side of Eq.~(\ref{eq:legendre}).

As in the standard approach to grand-canonical twist averaging, we start
by choosing a simulation cell and setting the chemical potential $\mu$.
We then calculate the particle numbers $N({\bf k}_s)$ and internal
energies $E({\bf k}_s)$ for $M$ different twists ${\bf k}_s$. The only
new feature is that we average the function of two independent
variables, $\Omega(\mu, N) = E(N) - \mu N$, instead of $E(N)$. Since
$\Omega(\mu, N)$ is stationary with respect to variations of $N$ about
the true particle number $N(\mu)$ at fixed $\mu$, the function
$\Omega(\mu, N)$ is relatively insensitive to small changes in $N$. The
values of $\Omega(\mu, N)$ obtained using different twists are therefore
good estimates of $\Omega(\mu)$, and the fluctuations in the
twist-averaged estimate of the grand potential,
\begin{equation}
  \label{eq:gcestimate}
  \Omega(\mu) = \frac{1}{M} \sum_{{\bf k}_s} \left ( E({\bf k}_s) - \mu
    N({\bf k}_s) \right ) ,
\end{equation}
are small.

Once this estimate of $\Omega(\mu)$ has been obtained, the internal
energy is easily found using the inverse Legendre transformation
\begin{equation}
  E = \Omega + \mu N ,
\end{equation}
where $\mu$ is the chosen chemical potential and $N$ is the expected
number of electrons in the simulation cell for that value of $\mu$. When
applied to a non-interacting system, this grand-potential
twist-averaging approach and the standard GCE twist-averaging approach
both yield the exact internal energy as the number of twists tends to
infinity, regardless of the size of the simulation cell. However, the
free-energy-averaging approach yields more accurate results when the
number of twists is small.

The chemical potential $\mu$ is known because it was chosen, but one
might expect the exact value of $N$ corresponding to a given $\mu$ to be
unknown in an interacting system. If this were the case, the inverse
Legendre transformation required to obtain $E$ from $\Omega$ could not
be carried out exactly in interacting systems. The most obvious solution
to this problem, which is to estimate $N$ via
\begin{equation}
  N = \frac{1}{M}\sum_{{\bf k}_s} N({\bf k}_s) ,
  \label{eq_N}
\end{equation}
is no good because the resulting internal energy estimate,
\begin{equation}
\begin{split}
        E &= \frac{1}{M} \sum_{{\bf k}_s} \left ( E({\bf k}_s) - \mu N({\bf k}_s)
\right ) + \frac{\mu}{M}\sum_{{\bf k}_s} N({\bf k}_s) \\
& = \frac{1}{M}\sum_{{\bf k}_s} E({\bf k}_s) ,
\end{split}
\end{equation}
reduces to Eq.\ (\ref{eq:Eave}), reintroducing the sensitivity to twist
and concomitant large fluctuations. Fortunately, in any practical
example, even for a correlated calculation, we \emph{do} know the mean
value of $N$ corresponding to any given $\mu$. The Slater determinant
part of the Slater-Jastrow trial function for a given twist contains
exactly the same number of electrons as the corresponding mean-field
wave function and the Jastrow factor does not change this. The mean value
of $N$, as obtained by an infinitely dense twist sampling, is thus
exactly the same as in the mean-field case and is easily calculated by
working out the volume of the mean-field Fermi surface.

The chosen chemical potential $\mu$, which is in practice obtained from
DFT or Hartree-Fock theory, will not be exactly equal to $dE/dN$ when
$E$ is the fully correlated energy. Consequently, $\Omega(\mu, N)$ will
not be exactly stationary with respect to small variations of $N$ about
its mean and the twist sensitivity of $\Omega(\mu, N)$ will be
increased. As long as the mean-field estimate of $\mu$ is reasonably
close to the true interacting chemical potential, however, the
fluctuations about the mean should still be much smaller than in the
internal-energy-based GCE twist-averaging approach. The
free-energy-based GCE twist-averaging algorithm therefore works almost
as well in fully correlated QMC simulations as in mean-field
calculations.

It is reassuring to note that the free-energy-based GCE twist-averaging
method yields exactly the same results as the internal-energy-based GCE
sampling method in the limit as the number of twists $M$ tends to
infinity, regardless of the accuracy of the estimate of $\mu$
employed. Averaging the free energy reduces the fluctuations but does
not affect the final estimate of the internal energy when the twist grid
is fine enough.

\section{Results}

\subsection{Uniform electron gas}
\label{subsec:UEG}

This section compares results obtained by applying three different
twist-averaging methods to the uniform electron gas with $r_s = 1$. The
energies were calculated in the mean-field Hartree-Fock approximation,
so twist averaging is here being used as an alternative to conventional
Brillouin-zone averaging of mean-field results. All calculations used a
Monkhorst-Pack grid of only $3 \times 3 \times 3$ twists (not all
inequivalent) centred on the $\Gamma$ point of the simulation-cell
Brillouin zone.

\begin{figure}
  \begin{center}
    \includegraphics[width=0.45\textwidth]{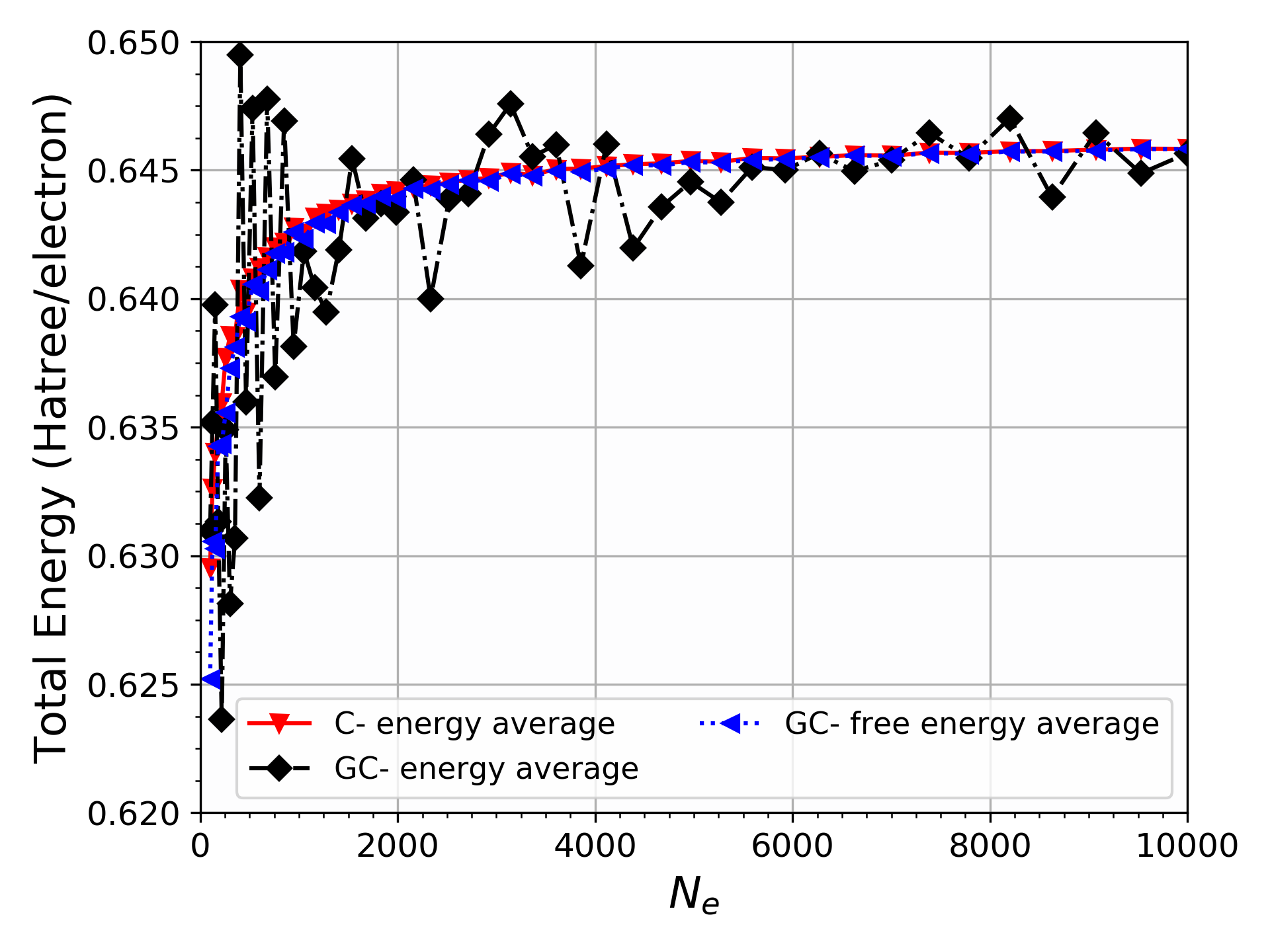}\\
    \vspace*{4mm}
    \caption{System-size dependence of the calculated total energy per
      electron of an $r_s = 1$ uniform electron gas in the Hartree-Fock
      approximation. Results obtained using canonical twist averaging of
      the total energy, grand-canonical twist averaging of the total
      energy, and grand-canonical twist averaging of the grand potential
      (free energy) are shown. In all cases, a $3 \times 3 \times 3$
      grid of twists centred on the $\Gamma$ point was
      used.\label{fig:E_potential_averaging}}
  \end{center}
\end{figure}

Figure \ref{fig:E_potential_averaging} shows that the ``random errors''
associated with the grand-canonical free-energy averaging algorithm are
much smaller than those associated with the grand-canonical internal
energy averaging algorithm and no larger than those associated with
canonical twist averaging of the internal energy. The systematic error
is dominated by the long-ranged Coulomb contribution to the exchange
energy, which cannot be removed by twist
averaging,\cite{Fraser,Drummond} but the additional small positive bias
caused by the approximation of the Fermi surface implicit in the
canonical twist-averaging algorithm can nevertheless be resolved.

\begin{figure}
  \begin{center}
    \includegraphics[width=0.45\textwidth]{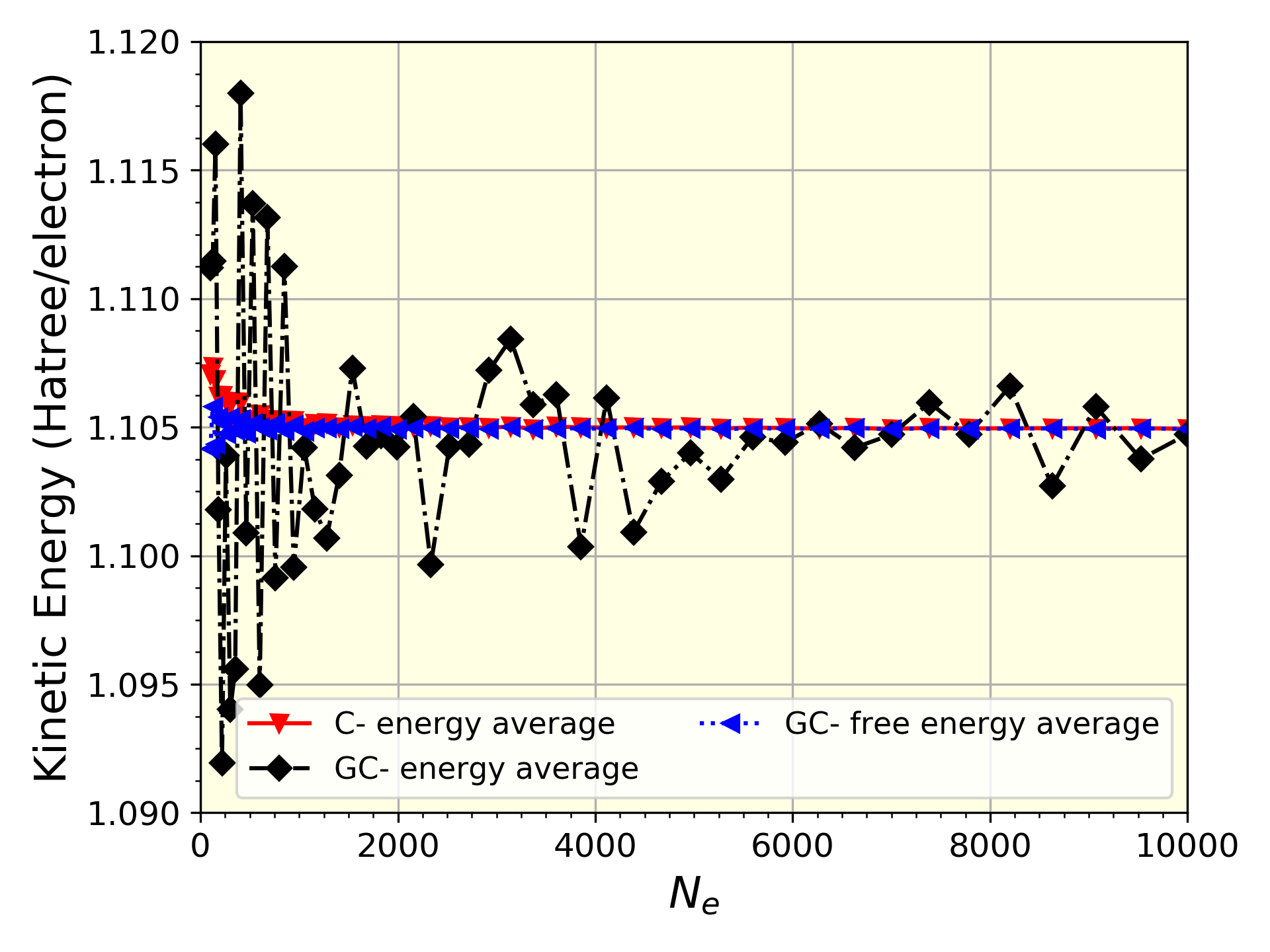}\\
    \vspace*{4mm}
    \caption{System-size dependence of the calculated kinetic energy per
      electron of an $r_s = 1$ uniform electron gas in the Hartree-Fock
      approximation. Results obtained using canonical twist averaging of
      the kinetic energy, grand-canonical twist averaging of the kinetic
      energy, and grand-canonical twist averaging of the kinetic
      component of the grand potential are shown. In all cases, a
      $3 \times 3 \times 3$ grid of twists centred on the $\Gamma$ point
      was used.\label{fig:KE_potential_averaging}}
  \end{center}
\end{figure}

\begin{figure}
  \begin{center}
    \includegraphics[width=0.45\textwidth]{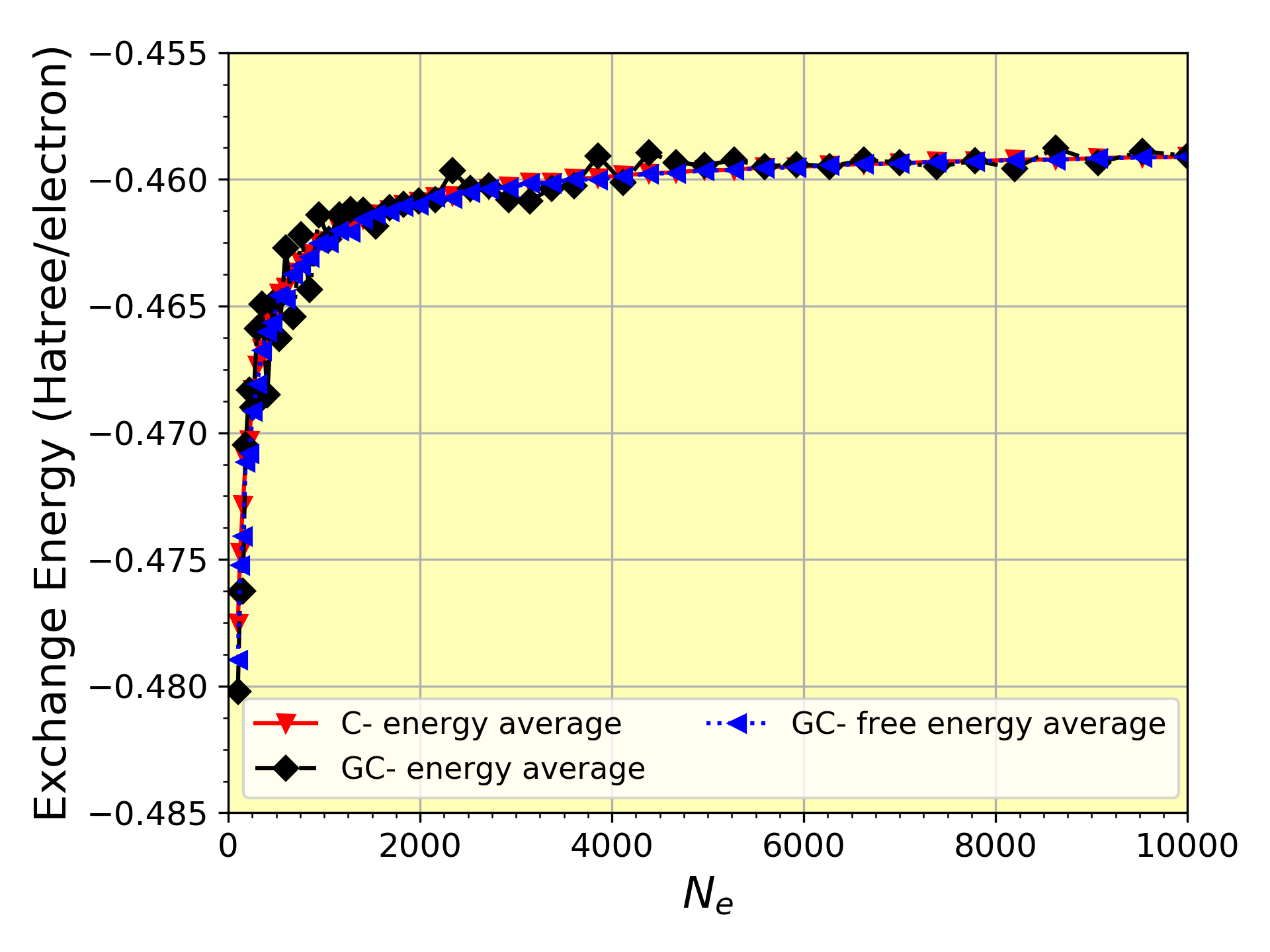}\\
    \vspace*{4mm}
    \caption{System-size dependence of the calculated exchange energy
      per electron of an $r_s = 1$ uniform electron gas in the
      Hartree-Fock approximation. Results obtained using canonical twist
      averaging of the exchange energy, grand-canonical twist averaging
      of the exchange energy, and grand-canonical twist averaging of the
      exchange component of the grand potential are shown. In all cases,
      a $3 \times 3 \times 3$ grid of twists centred on the $\Gamma$
      point was used.\label{fig:Ex_potential_averaging}}
  \end{center}
\end{figure}

As can be seen in Figs.\ \ref{fig:KE_potential_averaging} and
\ref{fig:Ex_potential_averaging}, the free-energy-based GCE
twist-averaging method works just as well for the kinetic energy, the
exchange energy, and presumably also other components of the total
energy. To obtain the kinetic and exchange energies, one averages the
kinetic and exchange components of the grand potential,
\begin{eqnarray}
\Omega_{T} & = & \frac{1}{M} \sum_{{\bf k}_s} \left ( T({\bf k}_s)
- \mu_{T}
N({\bf k}_s)\right ) , \\
\Omega_{E_{x}} & = & \frac{1}{M} \sum_{{\bf k}_s} \left ( E_{x}({\bf k}_s) -
\mu_{E_{x}} N({\bf k}_s) \right ) ,
\end{eqnarray}
where $T({\bf k}_s)$ is the kinetic energy of the simulation cell with
twist ${\bf k}_s$, $\mu_{T}$ is the kinetic contribution to the chemical
potential, $E_{x}({\bf k}_s)$ is the exchange energy of the simulation
cell with twist ${\bf k}_s$, and $\mu_{E_{x}}$ is the exchange
contribution to the chemical potential. For the Hartree-Fock
free-electron gas calculations carried out here, $\mu_{T}$ and
$\mu_{E_{x}}$ are given (in Hartree atomic units) by
$\mu_{T} = \frac{1}{2} k_f^2$ and $\mu_{E_{x}} = - \frac{1}{\pi} k_f$.

\subsection{Real metallic systems}
\label{subsec:metals}

This section investigates the value of grand-canonical grand-potential
twist averaging in DMC simulations of real metals.

The DMC calculations were carried out using the CASINO QMC package
\cite{casino} with Slater-Jastrow trial wave functions. The one-electron
orbitals appearing in the Slater determinants were generated within DFT
using the Quantum Espresso plane-wave code \cite{QE} with Trail-Needs
Dirac-Fock pseudopotentials.\cite{TN,NeilPRB} The Perdew-Burke-Ernzerhof
(PBE) generalized gradient approximation exchange-correlation functional
\cite {PBE} was used, and the plane-wave cutoff energy was set to 400 Ry
to obtain results close to the complete basis-set limit.\cite{samPRB10}
For Brillouin-zone integrations in metallic systems, we used the
Gaussian smearing scheme with the spreading parameter set to 25 meV. The
plane-wave representations of the one-electron orbitals were transformed
into a blip polynomial basis,\cite{blip} which is faster to evaluate in
QMC simulations. The Jastrow function consisted of polynomial one-body
electron-nucleus and two-body electron-electron terms, the parameters of
which were optimized by variance minimization at the variational Monte
Carlo level.\cite{variation_C,variation_N} We found the effect of
re-optimizing the Jastrow correlation function for every different twist
to be negligible, so the same optimized Jastrow function was used for
all twists. In all DMC calculations a time step of $\tau=0.005$ Hartree
atomic units of time was used.

Unlike the twists $\mathbf{k}_s$ used to obtain the electron gas results
desscribed in Sec.~\ref{subsec:UEG}, which were on a uniform
Monkhorst-Pack\cite{MPgrid} grid within the simulation-cell Brillouin
zone, the twists used for the QMC simulations of real materials reported
here were chosen randomly. Since the twists are chosen randomly, the
twist-dependent changes in the total energy may be treated as random
variables. The chemical potential $\mu$ was estimated from DFT
calculations. To ensure that the estimate of the DFT Fermi energy was
accurate, a dense $24 \times 24 \times 24$ ${\bf k}$-point mesh was
used.

When applying grand-canonical twist averaging (gctav) to real metallic systems
at zero temperature, we set the chemical potential $\mu$ to the
single-particle Fermi energy of the infinite system. As explained above,
errors in the value of $\mu$ increase the twist-dependent fluctuations
in the grand potential but do not affect the twist-averaged energy, so
the small difference between our choice of $\mu$ and the true
interacting chemical potential is unimportant. 
We applied the grand-potential twist-averaging method to three metallic
solids: high-pressure atomic hydrogen (H) in the tetragonal crystal
structure with $I4_1/amd$ symmetry;\cite{I4} lithium (Li) in the FCC
structure;\cite{Ackland} and FCC aluminum (Al). The numbers of atoms in
the simulation cells employed for the H, Li, and Al simulations were
128, 128, and 96, respectively. We used 16 random twists for H and Al,
and 18 random twists for Li.

Various exotic predictions have been made for atomic metallic hydrogen,
such as stability in a superfluid state or as a room-temperature
superconductor.\cite{Babaev,McMahon} Calculation of the phase diagram of
hydrogen and its electronic structure under extreme conditions is a
challenging subject for first-principles methods, not least because the
results obtained using DFT are severely affected by the choice of
exchange-correlation functional.\cite{PRB13, PRB19} The limitations of
DFT make DMC simulations of solid hydrogen particularly valuable, but
the accuracy required is very high and controlling the DMC finite-size
corrections is an important issue. This is particularly the case when
DMC is used to investigate the phase diagram.

\begin{figure}
  \includegraphics[width=0.45\textwidth]{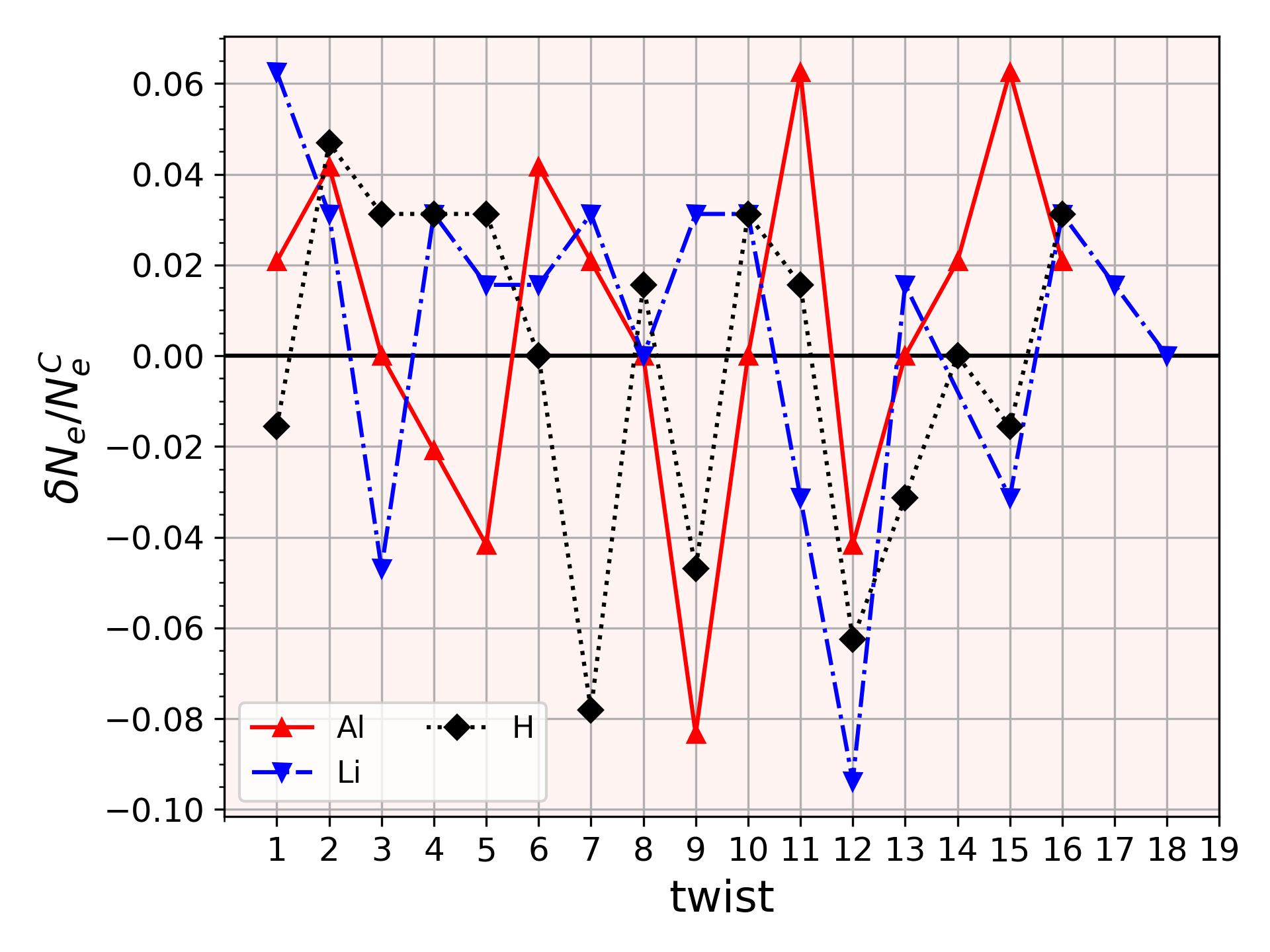}
  \caption{Relative fluctuations in the number of electrons in the
    grand-canonical simulation cell:
    $\delta N_{\text{e}} = N_{\text{e}}^{\text{C}} -
    N_{\text{e}}^{\text{GC}}$, where $N_{\text{e}}^{\text{C}}$ is the
    number of electrons occupying the simulation cell in the canonical
    ensemble and $N_{\text{e}}^{\text{GC}}$ is the number in the
    grand-canonical ensemble.\label{deltaNe}}
\end{figure}
Figure \ref{deltaNe} shows the relative fluctuations in the number of
electrons in the grand-canonical simulation cell as a function of twist
vector. The numbers of electrons per atom averaged over the 16 random
twists for H and Al and 18 random twists for Li may be evaluated as in
Eq.~(\ref{eq_N}). The results are 2.98(2), 0.99(2), and 1.00(1), for Al,
Li, and H, respectively. As the number of twists increases, the average
number of electrons per atom converges to the number of valence
electrons per atom as specified by the pseudopotential.

\begin{figure}
  \includegraphics[width=0.45\textwidth]{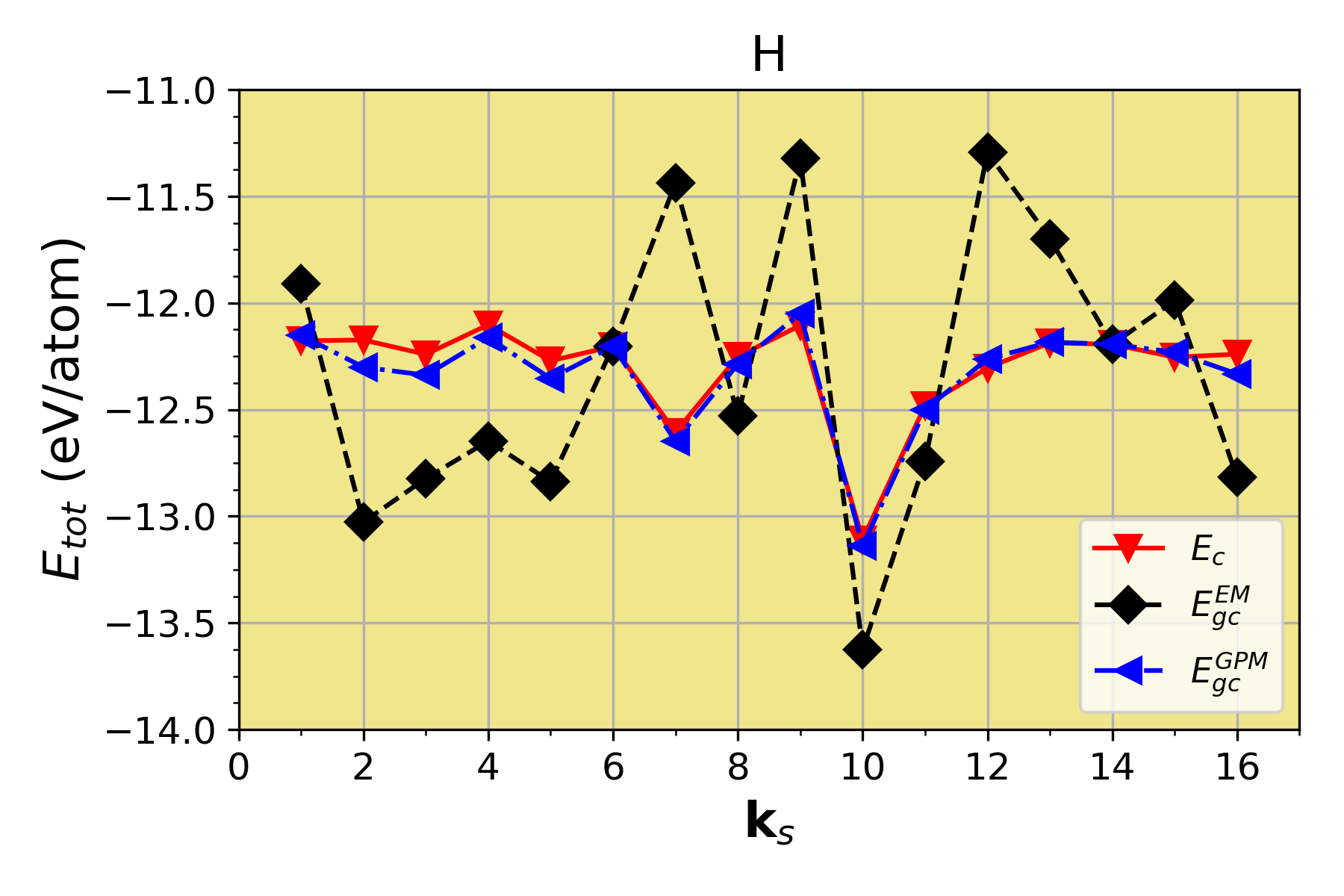}
  \includegraphics[width=0.45\textwidth]{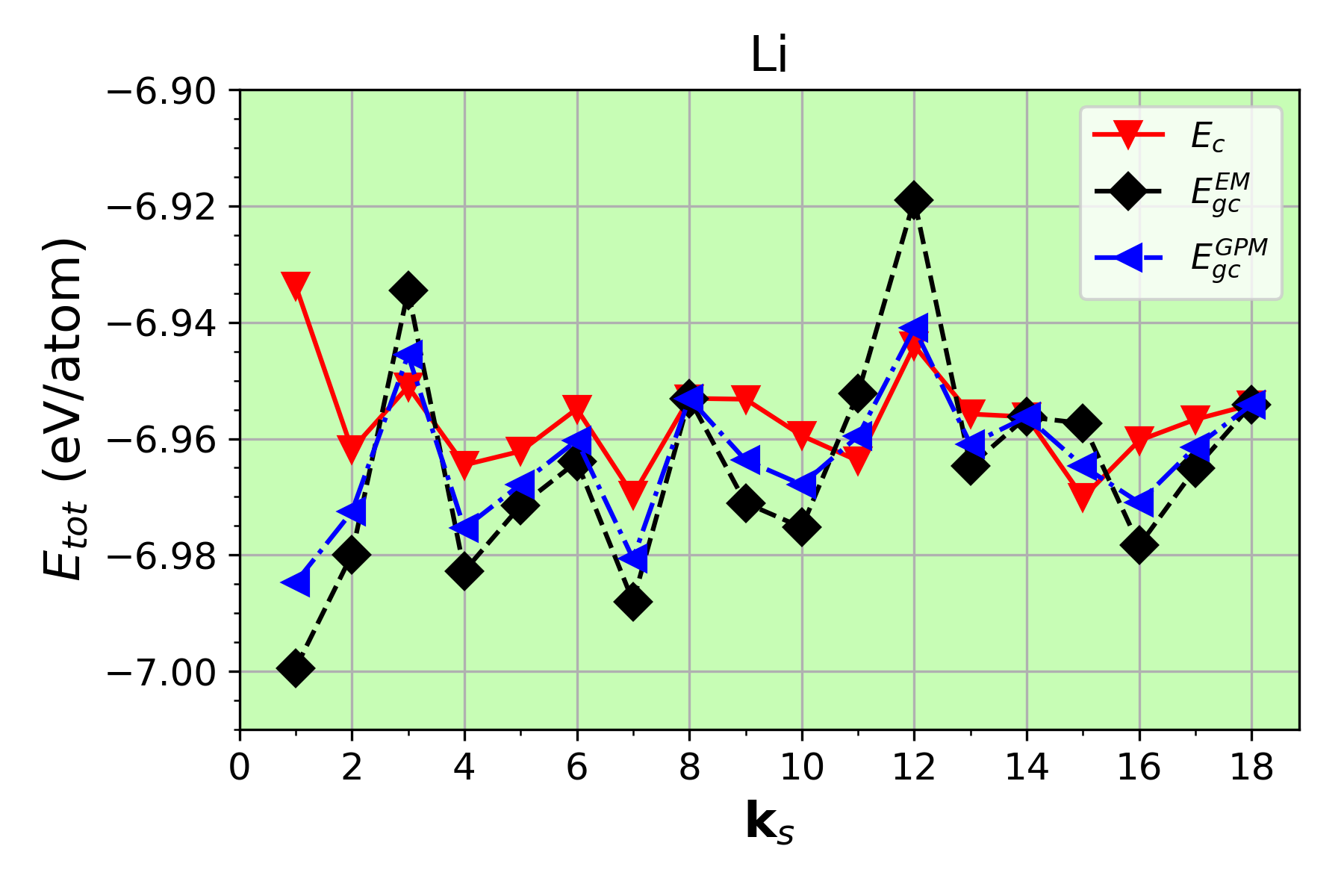}
  \includegraphics[width=0.45\textwidth]{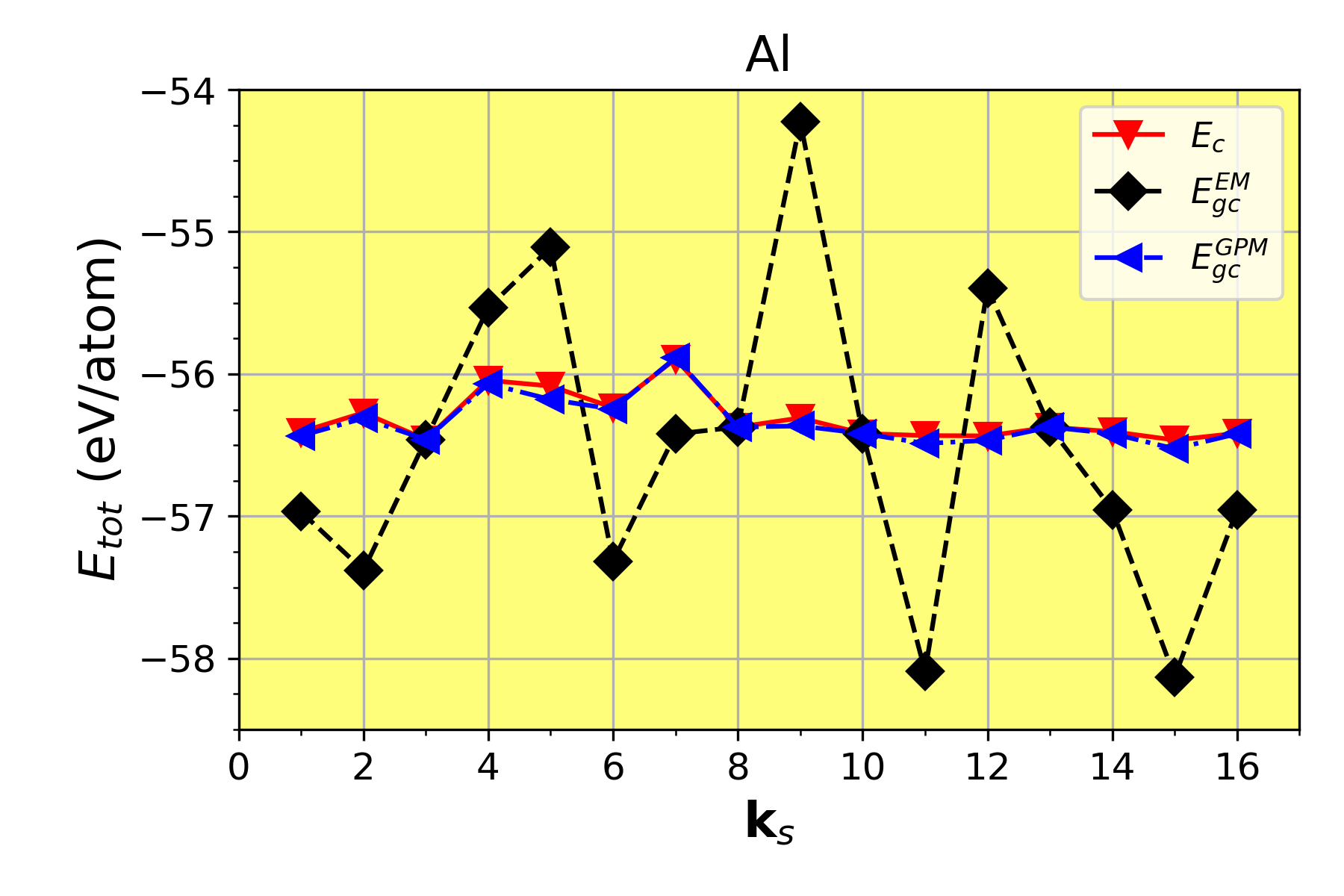}
  \caption{Twist dependence of the total DMC energy per atom for
    metallic H in the $I4_1/amd$ structure, FCC Li and FCC Al.
    The red triangles are internal energies ($E_c$)
    calculated using canonical simulations in which the number of
    electrons in the simulation cell is fixed. The black
    diamonds are internal energies ($E_{gc}^{EM}$) calculated
    using grand-canonical simulations in which the number $N_e$ of
    electrons in the simulation cell depends on the twist ${\bf k}_s$.
    The blue triangles are energies which are calculated by 
    $E_{gc}^{GPM} = \Omega({\bf k}_s, N_s) + \mu <N_e>$
    where $\Omega({\bf k}_s, N_s)$ is the grand canonical potential 
    defined as $E_{gc}^{EM} - \mu N_e({\bf k}_s)$, and $<N_e>$
    is the averaged number of electrons. 
    The statistical errors in all data points are smaller than the
    symbols.\label{ET}}
\end{figure}
Figure \ref{ET} shows our DMC results for metallic H, Li and Al. The
horizontal axis indexes the twists used, and the vertical axis shows the
total internal energy per atom for that twist. The red triangles,
$E_c({\bf k}_s)$, are energies calculated in the canonical ensemble,
with the number of electrons in the simulation cell fixed. The black
diamonds, $E_{gc}^{EM}({\bf k}_s)$, are energies calculated in the
grand-canonical ensemble, with the number of electrons in the simulation
cell dependent on the twist vector. The superscript $EM$ stands for
``energy method'', indicating that these results were not obtained using
the grand potential. As expected, the grand-canonical energy per atom is
considerably more sensitive to the twist than the canonical energy per
atom.

The blue triangles in Fig.~\ref{ET} are energies calculated using the
grand potential method ($GPM$):
\begin{equation}
  E_{gc}^{GPM}({\bf k}_s) = \Omega({\bf k}_s, N({\bf k}_s)) + \mu \langle N \rangle,
\end{equation}
where
\begin{equation}
  \Omega({\bf k}_s, N({\bf k}_s)) = E_{gc}^{EM}({\bf k}_s) - \mu
  N({\bf k}_s)
\end{equation}
is the estimate of the grand potential per atom at twist ${\bf k}_s$ and
$\langle N \rangle$ is the average number of electrons per atom as
defined by the pseudopotential. The standard deviation of $E_{gc}^{GPM}$
is much smaller than that of $E_{gc}^{EM}$ for all of the cases studied,
but especially for Al, which has a larger number of electrons in the
simulation cell.

The twist-averaged DMC energies for each system are reported in Table
\ref{results}. In all three metals the grand-canonical twist-averaged
energy lies below the canonical twist-averaged energy. Because all three
twist-averaging methods made use of the same random set of twists, the
statistical errors in energy differences are likely to be considerably
smaller than those in total energies.
\begin{table}
\begin{tabular}{|c c c c|}
\hline\hline
System & E$_{c}$ & E$_{gc}^{EM}$ & E$_{gc}^{GPM}$ \\
\hline
H & -12.31(6) & -12.3(2) & -12.33(6)\\
\hline
Li & -6.957(2) & -6.965(5) & -6.964(3)\\
\hline
Al & -56.31(4) & -56.5(3) &  -56.51(4) \\
\hline\hline
\end{tabular}
\caption{\label{results} Total energies in eV/atom of metallic H, Li,
  and Al obtained using canonical twist averaging (E$_{c}$),
  grand-canonical twist averaging of the internal energy
  (E$_{gc}^{EM}$), and grand-canonical twist averaging of the grand
  potential (E$_{gc}^{GPM}$). When working in the grand canonical
  ensemble, twist averaging the grand potential is much more efficient
  than twist averaging the internal energy.}
\end{table}
\section{Conclusion}
\label{conclude}

This paper presented a simple but efficient approach to twist averaging
in the grand-canonical ensemble. We explained that it is better to
average the grand potential $\Omega(\mu)$ than the internal energy. Once
the average of the grand potential has been found, the internal energy
can be obtained via a Legendre transformation,
$E(N) = \Omega(\mu) + \mu N$, where $\mu$ is the chosen chemical
potential and $N$ is the exact number of electrons per simulation cell.
Unlike conventional grand-canonical twist averaging of the internal
energy, the grand potential approach does not require very large numbers
of twists to provide accurate total energies; and unlike conventional
canonical twist averaging, the results are not biased when the
simulation cell is small. This makes grand-potential twist averaging in
the grand-canonical ensemble suitable for use in simulations of real
metallic systems, where the computational cost is a crucial factor.

\end{document}